\documentclass{article}

\usepackage{PRIMEarxiv}

\usepackage[utf8]{inputenc} 
\usepackage[T1]{fontenc}    
\usepackage{hyperref}       
\usepackage{url}            
\usepackage{booktabs}       
\usepackage{amsfonts}       
\usepackage{nicefrac}       
\usepackage{microtype}      
\usepackage{lipsum}
\usepackage{graphicx}
\graphicspath{{media/}}     

\usepackage{float}

\usepackage{tabularx}
\usepackage[table]{xcolor}
\usepackage{geometry}
\usepackage{listings}
\usepackage{xcolor} 

\lstdefinelanguage{SQL}{
  morekeywords={SELECT, FROM, WHERE, GROUP, BY, ORDER, AS, AVG, TO_CHAR},
  sensitive=false,
  morecomment=[l]{--},
  morestring=[b]',
}

\title{Uncovering Bottlenecks and Optimizing Scientific Lab Workflows with Cycle Time Reduction Agents
}

\author{
  Yao Fehlis \\
  Artificial, Inc. \\
  \texttt{yao@artificial.com} 
}

\begin{document}
\maketitle

\begin{abstract}
Scientific laboratories, particularly those in pharmaceutical and biotechnology companies, encounter significant challenges in optimizing workflows due to the complexity and volume of tasks such as compound screening and assay execution. We introduce Cycle Time Reduction Agents (CTRA), a LangGraph-based agentic workflow designed to automate the analysis of lab operational metrics. CTRA comprises three main components: the Question Creation Agent for initiating analysis, Operational Metrics Agents for data extraction and validation, and Insights Agents for reporting and visualization, identifying bottlenecks in lab processes. This paper details CTRA’s architecture, evaluates its performance on a lab dataset, and discusses its potential to accelerate pharmaceutical and biotechnological development. CTRA offers a scalable framework for reducing cycle times in scientific labs.
\end{abstract}


\section{Introduction}
Scientific laboratories in pharmaceutical and biotechnology companies manage intricate workflows involving thousands of tasks daily, encompassing sample preparation, compound screening, assay execution, and data analysis. These tasks are meticulously tracked in databases with metrics such as creation timestamps, completion statuses, and execution durations. Identifying inefficiencies—such as prolonged screening processes or elevated assay failure rates—is paramount to accelerating pharmaceutical and biotechnological development and minimizing time-to-market. However, traditional methods for bottleneck analysis rely heavily on manual processes, where scientists spend hours crafting queries, analyzing data, and creating visualizations. This labor-intensive approach is not scalable and diverts valuable time from core research activities, highlighting the need for automated solutions to streamline lab operations~\cite{fehlis2025acceleratingdrugdiscoveryartificial}.

Recent advances in LLMs have significantly impacted scientific research, particularly in biotechnology, chemistry, materials science, and drug discovery. LLMs such as BERT, GPT, and specialized models like ChemBERTa~\cite{chithrananda2020chemberta} have been fine-tuned to understand domain-specific languages, enabling applications such as protein sequence prediction~\cite{xiao2024proteingpt,Dodero-Rojas2024.07.17.603864}, chemical reaction modeling~\cite{chen2025reactgpt}, and materials property prediction~\cite{li2023call,tang2025matterchat,wang2025evaluatingperformancerobustnessllms}. In drug discovery, LLMs have accelerated tasks such as hypothesis generation, molecular property prediction, and literature-based target identification by enabling context-aware reasoning over large, heterogeneous biomedical datasets. Their integration into automated workflows has shown promise in improving lead optimization and reducing experimental cycles~\cite{irwin2022chemformer,luo2022biogpt,song2025llm,brahmavar2024generating}. Furthermore, the emergence of self-driving laboratories — autonomous systems that integrate AI with robotic experimentation — has revolutionized experimental workflows~\cite{tom2024self,hysmith2024future,abolhasani2023rise}. These labs leverage LLMs to design experiments, interpret results, and iteratively optimize processes in real-time, as demonstrated in applications like automated synthesis of organic compounds and high-throughput materials discovery. The integration of LLMs into self-driving labs has enabled unprecedented efficiency in scientific discovery, reducing human intervention and accelerating innovation across pharmaceutical and biotechnological sectors~\cite{zou2025elagenteautonomousagent}. 

Agentic AI, which builds on LLMs by adding autonomous decision-making and multi-step reasoning capabilities, has gained traction in scientific domains like biotechnology, chemistry, and materials science. In biotech, agentic systems have improved high-throughput screening and real-time process monitoring, addressing challenges in pharmaceutical lab operations. In chemistry, they have optimized synthesis pathways, while in materials science, they support self-driving labs for high-throughput discovery~\cite{inoue2025drugagent,huang2024protchat,ghafarollahi2024atomagentsalloydesigndiscovery,schmidgall2025agent}. Agentic AI leverages frameworks like LangGraph to coordinate complex workflows, enabling applications such as experiment automation in self-driving labs and data-driven optimization in lab settings. These advancements demonstrate agentic AI’s potential to tackle data-intensive challenges in scientific research.

We propose Cycle Time Reduction Agents (CTRA), a LangGraph-based workflow that automates lab analytics with a Question Creation Agent, Operational Metrics Agents, and Insights Agents. CTRA targets bottleneck identification by analyzing metrics such as execution times and error rates, enabling pharmaceutical and biotechnology companies to optimize workflows. For example, it can detect delays in compound screening or assay failures, offering data-driven recommendations. This paper contributes a novel agentic workflow for lab analytics, an evaluation framework with detailed results, and insights into its practical applications in scientific lab automation.

\section{Method}
\label{sec:method}

\subsection{Problem Formulation}
Given a lab database with fields (e.g., \textit{created\_timestamp}, \textit{state}), the goal is to identify bottlenecks by automatically creating questions (e.g., “What is the average duration of failed assays?”), answering them with SQL queries and proposing human-actionable suggestions and insights. CTRA automates this with robustness and scalability across three components: Question Creation Agent, Operational Metrics Agents, and Insights Agents.

\subsection{System Architecture}
CTRA uses a LangGraph workflow with three main components: the Question Creation Agent, Operational Metrics Agents, and Insights Agents. 

\begin{figure}[H]
  \centering
  \includegraphics[width=1.0\textwidth]{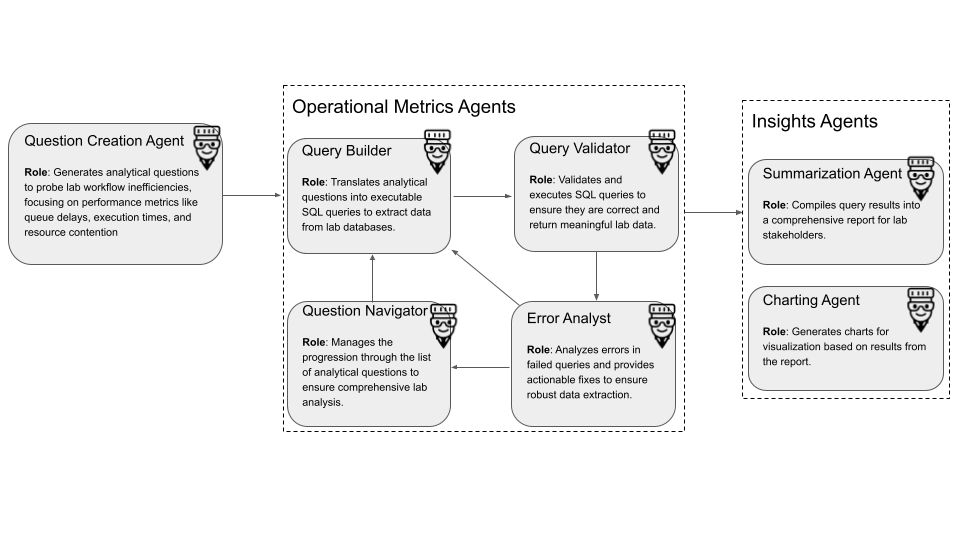} 
  \caption{Overview of the Cycle Time Reduction Agents (CTRA) workflow for scientific lab optimization. The process begins with the Question Creation Agent generating analytical questions, followed by Operational Metrics Agents (Query Builder, Query Validator, Error Analyst, Question Navigator) extracting and refining data, and Insights Agents (Summarization Agent, Charting Agent) compiling reports and visualizations. Conditional routing based on query outcomes reflects CTRA’s dynamic decision-making for lab analytics.}
  \label{fig:fig1}
\end{figure}

\subsubsection{CTRA Roles and Responsibilities}
CTRA comprises seven agents, organized into three main components:

\textbf{Question Creation Agent} 
\begin{enumerate}
 \item \textbf{Question Creation Agent}:
 \begin{itemize}
  \item Role: Generates analytical questions to probe lab inefficiencies, focusing on metrics like queue delays and execution times.
  \item How It Works: Powered by large language models (LLMs), it creates questions tailored to lab metrics, such as "What is the daily count of new compound tests? (Suitable for line chart)" or "What is the average duration of failed assays? (Suitable for bar chart)". This ensures relevance to pharmaceutical workflows.
  \item Example: ["What is the daily count of new compound screening tasks? (Suitable for line chart)"].
  \end{itemize}
\end{enumerate}

\textbf{Operational Metrics Agents} 
\begin{enumerate}
 \item \textbf{Query Builder}:
 \begin{itemize}
  \item Role: Translates questions into SQL queries for lab data extraction.
  \item How It Works: Leveraging code generation LLMs, it produces SQL queries based on the question and lab schema, with retry logic to refine queries if errors occur.
  \item Retry Logic: Uses error feedback to correct queries, ensuring accurate lab data retrieval.
  \item Example: For the question "What is the daily average execution time of jobs?", the agent produces:
\begin{lstlisting}
SELECT TO_CHAR(execution_timestamp, 'YYYY-MM-DD') AS date,
       AVG(execution_time) AS avg_execution_time
FROM jobs
GROUP BY TO_CHAR(execution_timestamp, 'YYYY-MM-DD')
ORDER BY date
\end{lstlisting}
  \end{itemize}

   \item \textbf{Query Validator}:
 \begin{itemize}
  \item Role: Validates and executes SQL queries to ensure they are correct and return meaningful lab data.
  \item How It Works: Checks queries for syntax and schema compliance, executes them using Artificial’s secure tools, and returns results or errors. 
  \item Example: For the query above, the agent validates the syntax, confirms the use of valid columns (\textit{execution\_timestamp}, derived from \textit{completed\_timestamp - started\_timestamp}), and ensures the output format (date, average time) is suitable for visualization.
  \end{itemize}

     \item \textbf{Question Navigator}:
 \begin{itemize}
  \item Role: Manages the progression through the list of analytical questions to ensure comprehensive lab analysis.
  \item How It Works: Advances to the next question after processing, resetting state, or triggers the Summarization Agent if all questions are done, ensuring seamless workflow navigation.
  \item Example: After processing "What is the daily average execution time of jobs?", the agent moves to the next question, such as "What is the average creation-to-start time for jobs, grouped by state?", ensuring all questions are systematically addressed.
  \end{itemize}

       \item \textbf{Error Analyst}:
 \begin{itemize}
  \item Role: Analyzes errors in failed queries and provides actionable fixes to ensure robust data extraction.
  \item How It Works: Proposes fixes for issues like "column 'status' must appear in GROUP BY," feeding suggestions back to the Query Builder, with up to three iterations to ensure lab data integrity.
  \item Example: For a failed query with error "column 'status' must appear in GROUP BY", the agent suggests adding \lstinline[language=SQL]{GROUP BY state}
 (correcting the column name) and retries the query execution.
  \end{itemize}
\end{enumerate}

\textbf{Insights Agents} 
\begin{enumerate}
 \item \textbf{Summarization Agent}:
 \begin{itemize}
  \item Role: Compiles query results into a comprehensive report for lab stakeholders.
  \item How It Works: Generates a report with insights like "Compound screening process X has the highest execution time". The report has human-actionable suggestions.
  \item Example: Given results showing one workflow with over 41,000 errors, the agent compiles a report noting: "Analysis revealed a workflow with 95\% of errors, indicating a critical bottleneck. Recommendations include protocol standardization and equipment checks."
  \end{itemize}

 \item \textbf{Charting Agent}:
 \begin{itemize}
  \item Role: Generates charts for visualization based on results from the report.
  \item How It Works: Creates charts (e.g., line charts for daily compound tests) using Matplotlib, aiding lab managers in interpreting data visually.
  \item Example: For "How many jobs have errors in logs, grouped by workflow\_id?", the agent generates a bar chart with workflow IDs on the x-axis and error counts on the y-axis.
  \end{itemize}
\end{enumerate}

\section{Experiments}
\label{sec:Experiments}
In this session, we describe in detail the experimental setup, including the dataset and the configuration of the employed large language models (LLMs) used in the agents. 

\subsection{Data}

The data is collected using the Artificial platform (artificial.com), a sophisticated tool designed to streamline and enhance the management of scientific experiments. When running experiments, the platform automatically records key metrics such as job states, execution records, logs, and timestamps within the PostgreSQL 'jobs' table. This data is gathered ahead of time, however, the same methodology works for real-time experiments as well. The platform's integration with lab workflows enables seamless data aggregation, providing a robust foundation for identifying bottlenecks and optimizing operational efficiency.

The dataset is sourced from a PostgreSQL table named \textit{jobs}, containing approximately 5,000 records with the following schema:

\rowcolors{2}{gray!10}{white}
\begin{tabularx}{\textwidth}{>{\raggedright\arraybackslash}p{4.5cm} >{\raggedright\arraybackslash}p{4.5cm} X}
\hline
\textbf{Column Name} & \textbf{Data Type} & \textbf{Description / Notes} \\
\hline
id & VARCHAR(255) & Primary key \\
name & VARCHAR(255) & Not null \\
lab\_id & VARCHAR(255) & Not null \\
workflow\_id & VARCHAR(255) & Not null \\
state & VARCHAR(50) & Not null (e.g., 'COMPLETED', 'IN\_ERROR') \\
created\_timestamp & TIMESTAMP WITH TIME ZONE & Not null \\
started\_timestamp & TIMESTAMP WITH TIME ZONE & Optional \\
completed\_timestamp & TIMESTAMP WITH TIME ZONE & Optional \\
root\_action\_id & VARCHAR(255) & Optional \\
lab\_reference & VARCHAR(255) & Optional \\
associated\_ids & JSONB & Optional \\
parameters & JSONB & Optional \\
outputs & JSONB & Optional \\
barcodes & JSONB & Optional \\
batched\_job\_ids & JSONB & Optional \\
children\_job\_ids & JSONB & Optional \\
execution\_records & JSONB & Array of objects with 'event\_type', 'name', 'started\_timestamp', 'finished\_timestamp' \\
logs & JSONB & Array of objects with 'level', 'message', 'created\_timestamp' \\
files & JSONB & Optional \\
notes & JSONB & Optional \\
configuration\_versions & JSONB & Optional \\
\hline
\end{tabularx}

The data represents a month of lab operations, including compound screening and assay execution tasks in a pharmaceutical and biotechnology setting. The dataset encompasses approximately 5,000 records, with data quality generally high, though occasional NULL values in fields like \textit{started\_timestamp} and \textit{execution\_records} required handling during query validation. 

\subsection{LLMs}

CTRA utilizes multiple large language models (LLMs) to support its agentic workflow, each selected for its specific strengths in natural language understanding, code generation, and analytical reasoning. For the Question Creation and Charting Agents, we employ LLaMA-3.1-70B-Instruct, a 70-billion-parameter model optimized for instruction-following tasks, known for its ability to generate coherent natural language questions and Python code for visualizations. The Query Builder leverages DeepSeek-R1~\cite{guo2025deepseek}, a specialized model designed for technical code generation, particularly excelling in producing accurate SQL queries for database operations. For the Summarization Agent, we use LLaMA-3.1-405B-Instruct, a larger 405-billion-parameter variant of LLaMA, which offers advanced reasoning capabilities for compiling detailed analytical reports. Each model is configured with a temperature of 0.7 to balance creativity and accuracy, and a maximum token limit of 4,000 to handle complex prompts. 

\section{Results}
CTRA has demonstrated significant potential in transforming lab operations for pharmaceutical and biotechnology companies by uncovering bottlenecks and providing actionable insights. Below, we detail the examples of CTRA, including the automatically generated queries, human-actionable suggestions, corresponding visualizations, and findings.

\begin{figure}[H]
  \centering
  \includegraphics[width=0.9\textwidth]{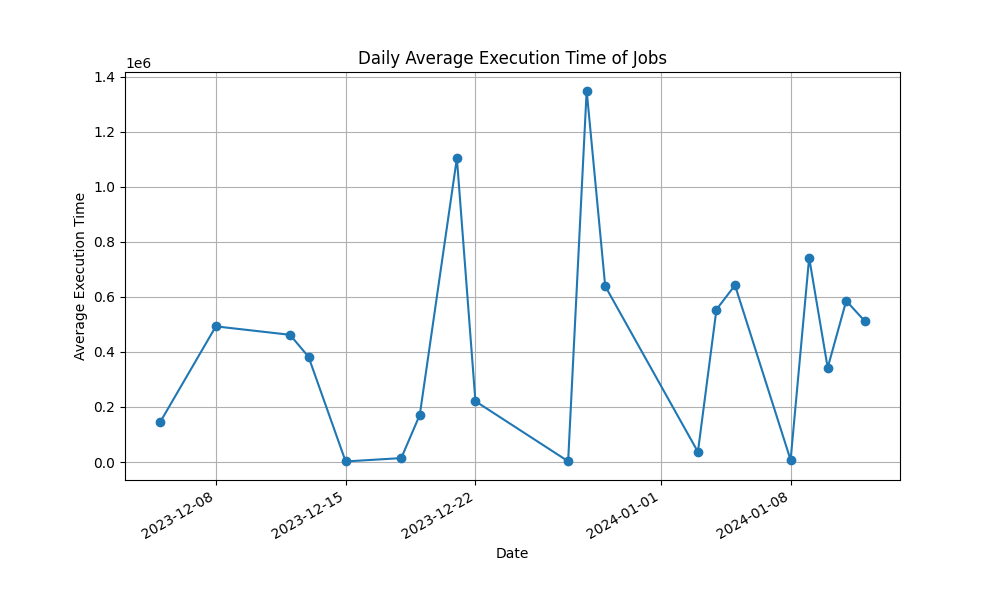} 
  \caption{Daily Average Execution Time of Jobs.}
  \label{fig:fig2}
\end{figure}

\textbf{Example 1:} 
 \begin{itemize}
  \item Automatically Generated Query:  "What is the daily average execution time of jobs?"
  \item Findings: The analysis revealed significant variability in daily execution times, with peaks exceeding 1.3 million seconds on some days and dropping to as low as 2,100 seconds on others. This fluctuation suggests potential inefficiencies in scheduling or resource allocation, which could delay pharmaceutical development processes. See Figure \ref{fig:fig2} for visualization. 
  \item Human-Actionable Suggestions:
  \begin{itemize}
  \item Optimize scheduling of lab jobs to reduce variability in execution times.
  \item Standardize protocols for lab jobs to minimize setup times and reduce errors.
  \item Implement a robust error handling mechanism to reduce error rates and minimize execution times.
  \item Regularly review and optimize lab job protocols to reduce setup times and improve efficiency.
  \item Consider upgrading equipment or software to improve processing times and reduce JSONB processing overhead.
  \end{itemize}
\end{itemize}

\begin{figure}[H]
  \centering
  \includegraphics[width=0.8\textwidth]{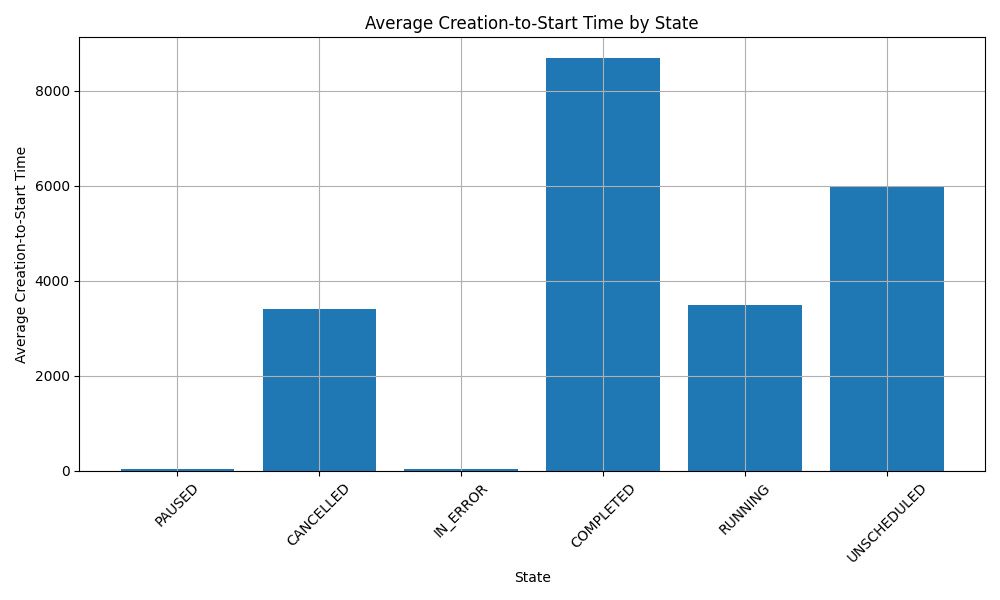} 
  \caption{Average Creation-to-Time by State.}
  \label{fig:fig3}
\end{figure}

\textbf{Example 2:} 
 \begin{itemize}
  \item Automatically Generated Query: "What is the average creation-to-start time for jobs, grouped by state?"
  \item Findings: The results showed significant variability, with completed jobs averaging over 8,600 seconds and cancelled jobs around 3,400 seconds, while paused jobs were much faster at 33 seconds. This indicates potential bottlenecks in the completion and cancellation processes, possibly due to manual interventions or system delays, impacting overall lab efficiency. See Figure \ref{fig:fig3} for visualization. 
  \item Human-Actionable Suggestions:
\begin{itemize}
  \item Optimize job scheduling to reduce creation-to-start times for jobs in the \texttt{COMPLETED} state.
  \item Improve cancellation procedures to minimize creation-to-start times for jobs in the \texttt{CANCELLED} state.
  \item Standardize protocols for job scheduling and completion to minimize variability in creation-to-start times.
  \item Regularly review and refine job cancellation procedures to ensure efficiency.
  \item Implement data quality checks to ensure accurate and consistent timestamp records.
\end{itemize}
\end{itemize}

\begin{figure}[H]
  \centering
  \includegraphics[width=0.8\textwidth]{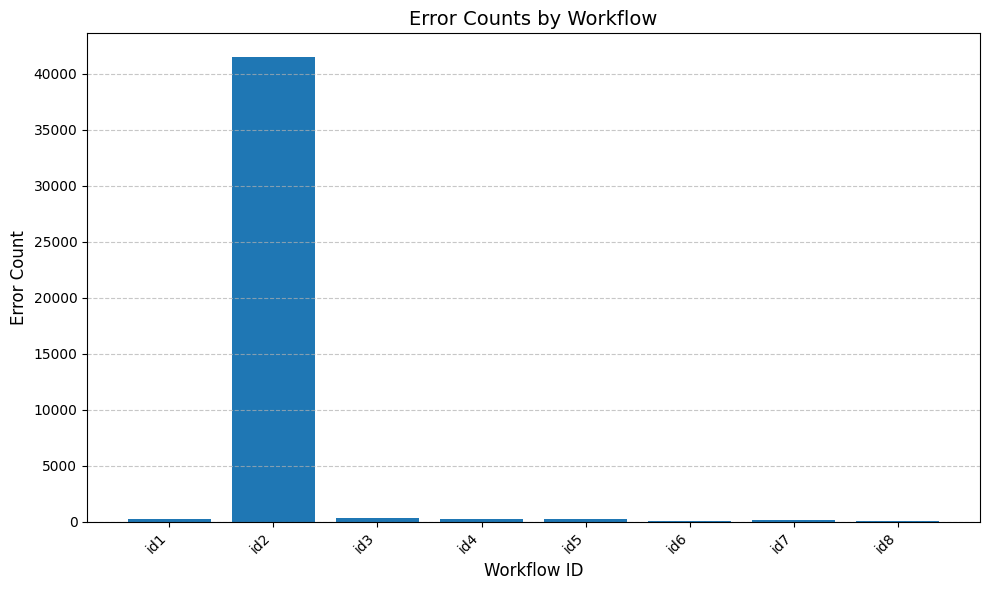} 
  \caption{Error Counts by Workflow.}
  \label{fig:fig4}
\end{figure}

\textbf{Example 3:} 
 \begin{itemize}
  \item Automatically Generated Query: "How many jobs have errors in logs, grouped by workflow\_id?"
  \item Findings: The analysis revealed that one workflow accounted for approximately 95\% of errors (over 41,000 errors), while others had significantly fewer errors (ranging from 10 to 330). This concentration suggests a critical bottleneck in that workflow, likely due to protocol inconsistencies or equipment issues, which could significantly hinder lab operations if not addressed. See Figure \ref{fig:fig4} for visualization. 
  \item Human-Actionable Suggestions:
\begin{itemize}
  \item Review and standardize protocols for the workflow with the highest errors to minimize errors and optimize performance.
  \item Check equipment and software used in the high-error workflow to ensure they are functioning correctly and not contributing to errors.
  \item Consider re-training personnel involved in the high-error workflow to ensure they are following best practices and protocols.
  \item Regularly review and standardize protocols for all workflows to minimize errors and optimize performance.
  \item Implement regular maintenance and quality control checks on equipment and software used in lab workflows.
  \item Consider investing in additional training for lab personnel to ensure they are following best practices and protocols.
\end{itemize}
\end{itemize}

\section{Summary and Future Directions}
CTRA automates lab analytics, offering a robust framework for optimizing pharmaceutical and biotechnological development by identifying bottlenecks in workflows. The system’s ability to generate actionable insights from complex datasets, as demonstrated by the analysis of execution times, creation-to-start delays, and error rates, highlights its practical utility. These human-actionable suggestions are particularly valuable for scientists and automation engineers. Scientists can utilize recommendations such as protocol standardization and personnel retraining to enhance experimental consistency and reduce errors, directly improving research outcomes. Automation engineers can leverage suggestions like equipment upgrades and scheduling optimization to design more efficient lab systems, ensuring seamless integration of automated processes. The integration of LangGraph and agentic AI enables scalable and repeatable analysis, making CTRA a valuable tool for scientific labs. The results suggest that targeted interventions, such as scheduling optimization and protocol standardization, can significantly enhance lab efficiency. This study lays the groundwork for broader adoption of agentic workflows in scientific research, with potential applications extending beyond pharmaceutical and biotechnology labs to other data-intensive domains.
Future enhancements include real-time monitoring, interactive dashboards, and extensions to other scientific domains.

\bibliographystyle{unsrt}  
\bibliography{references}

\appendix
\section{Appendix}

\subsection{Prompts for Agentic Workflow}
This subsection presents the system prompts used to configure the agents in the CTRA workflow. These prompts are designed to guide the agents in generating questions, SQL queries, error analysis, reports, and visualizations, ensuring alignment with the goals of bottleneck identification and lab optimization. Each prompt is presented in a boxed format for clarity.

\subsubsection{GENERATE\_QUESTIONS\_PROMPT}

\begin{lstlisting}
You are an expert in PostgreSQL performance optimization. Below is the schema of a PostgreSQL database table:

{table_schema}

Generate exactly {num_questions} human-language questions to identify potential operational bottlenecks in the 'jobs' table. Each question must focus on comparative analysis across categories, time intervals, or other groupings (e.g., comparing job counts, execution times, or JSONB-derived metrics across workflow_id, lab_id, or daily/weekly periods) to highlight performance differences. Questions must avoid single-value results (e.g., total job count) and instead produce results with multiple values for comparison. Each question should be phrased in natural language, designed to be translated into SQL queries to analyze performance issues such as slow queries, inefficient indexing, high JSONB processing overhead, or concurrency problems. Leverage columns like 'state', 'workflow_id', 'execution_records', and 'logs', and include safe JSONB operations when appropriate. Return a JSON array of strings, where each string is a human-readable question. Return only the JSON array.

**Instructions**:
- Focus on performance metrics (e.g., execution time, job counts, JSONB-derived durations) that allow comparison across groups or time periods.
- Use only columns from the schema: id, name, lab_id, workflow_id, state, created_timestamp, started_timestamp, completed_timestamp, root_action_id, lab_reference, associated_ids, parameters, outputs, barcodes, batched_job_ids, children_job_ids, execution_records, logs, files, notes, configuration_versions.
- For JSONB fields ('execution_records', 'logs'), allow simple key access using `->` or `->>` for top-level keys (e.g., 'status', 'duration', 'error_message'). Assume these keys exist and return text or numeric values. Use COALESCE to handle missing keys (e.g., COALESCE(execution_records->>'duration', '0')::FLOAT). Avoid complex JSONB operations (e.g., nested key access, array iteration, pattern matching).
- Ensure questions produce **2-3 columns** (e.g., category, value) for visualization (bar charts, line graphs).
- Prefer questions that:
  - Group by categorical columns (e.g., workflow_id, lab_id, state) or time intervals (e.g., daily, weekly).
  - Return counts, averages, or other aggregates for comparison across groups or time periods.
  - Explicitly avoid single-value results (e.g., total job count or average execution time for all jobs).
  - Include JSONB-derived metrics (e.g., average duration from execution_records) when relevant, ensuring safe key access.
- Avoid questions assuming specific states (e.g., 'IN_ERROR') unless verified.
- Include visualization suggestions (e.g., 'Suitable for bar chart').

Example output:
[
    "What is the daily average execution time of jobs, grouped by workflow_id? (Suitable for line chart)",
    "How do job counts compare across different lab_id values? (Suitable for bar chart)",
    "What is the average creation-to-start time for jobs, grouped by state? (Suitable for bar chart)",
    "What is the average duration from execution_records, grouped by lab_id? (Suitable for bar chart)",
    "How many jobs have errors in logs, grouped by workflow_id? (Suitable for bar chart)"
]
\end{lstlisting}
\normalsize

\subsubsection{GENERATE\_SQL\_PROMPT}
\begin{lstlisting}
Given the following PostgreSQL table schema:
{table_schema}

Generate a raw PostgreSQL SELECT query to answer the following question:
{question}

**Instructions**:
- Output **only** a single, executable PostgreSQL SELECT query as plain text.
- The query **must** start with 'SELECT' and be a top-level SELECT statement (no CTEs, subqueries, or WITH clauses).
- Use **only** columns defined in the schema.
- Ensure the query is valid, executable PostgreSQL.
- Keep queries simple: avoid window functions, complex JSONB operations, or nested queries.
- Return **2-3 columns** (e.g., category, value) for visualization.
- Convert timestamps to strings using TO_CHAR (e.g., TO_CHAR(created_timestamp, 'YYYY-MM-DD')).
- For time calculations, use EXTRACT(EPOCH FROM (timestamp2 - timestamp1)) and filter NULL timestamps.
- Use COALESCE for aggregates (e.g., COALESCE(AVG(...), 0)).
- Avoid hard-coded filters (e.g., state = 'IN_ERROR') unless required.
- **Do not** include comments, empty lines, or extra whitespace.
- Ensure the query produces visualization-friendly output (e.g., categorical or time-based columns with numeric values).

**Example**:
Question: "How many jobs are processed per workflow_id?"
Query:
SELECT workflow_id, COUNT(*) AS job_count FROM jobs GROUP BY workflow_id ORDER BY job_count DESC

**Output**:
A single, executable PostgreSQL SELECT query.
\end{lstlisting}

\subsubsection{CODE\_CHECK\_PROMPT}
\begin{lstlisting}
Given the following PostgreSQL query:
{code}

And the question it aims to answer:
{question}

Check the query for:
- Syntax errors (e.g., invalid PostgreSQL syntax).
- Logical errors (e.g., incorrect joins, aggregations, or filters).
- Performance issues (e.g., missing indexes, inefficient subqueries).
- Alignment with the question (e.g., does it answer the question correctly?).
- Use of only valid columns from the schema: id, name, lab_id, workflow_id, state, created_timestamp, started_timestamp, completed_timestamp, root_action_id, lab_reference, associated_ids, parameters, outputs, barcodes, batched_job_ids, children_job_ids, execution_records, logs, files, notes, configuration_versions.

**Instructions**:
- Output **only** a valid JSON object. Do **not** include markdown (e.g., ```json), explanatory text, or additional SQL queries.
- Ensure the response contains:
  - is_valid: boolean (true if the query is valid and answers the question, false otherwise).
  - errors: list of strings (describe any issues found).
  - suggestions: list of strings (provide optimization or correction suggestions).

**Example Response**:
{
    "is_valid": false,
    "errors": ["Invalid column 'invalid_col'", "Missing GROUP BY for non-aggregated column"],
    "suggestions": ["Use valid column 'workflow_id'", "Add GROUP BY workflow_id"]
}
\end{lstlisting}

\subsubsection{REFLECT\_PROMPT}
\begin{lstlisting}
Given the PostgreSQL query:
{code}
And the error messages:
{errors}
Reflect on how to fix the issues and improve the query for the jobs database:
{table_schema}
Provide a brief explanation of the changes needed.
\end{lstlisting}

\subsection{REPORT\_PROMPT}
\begin{lstlisting}
You are an expert data analyst tasked with writing a comprehensive, human-readable report summarizing operational bottlenecks in a PostgreSQL 'jobs' table. The table schema is:

{table_schema}

Below are the queries and their results:

{queries_results}

For each query, use the provided results (columns and data) to write a detailed report with the following structure:

- Introduction: Describe the purpose of the report and the context of the 'jobs' table (e.g., it tracks job workflows with states, execution records, logs, and timestamps).
- Analysis: For each query, provide:
The query question.
- Results in a clear format (e.g., table of columns and data).
- Bottleneck insights, explaining how the results indicate performance issues (e.g., high error rates, JSONB query complexity).
- Reference a visualization saved as '{output_dir}/plot_query_X.png' if a plot was generated, or note if no plot was generated due to errors.
Recommendations: List 5 actionable recommendations to address the bottlenecks, written in a professional tone.
- Conclusion: Summarize the key bottlenecks and the importance of addressing them.

Rules:

- Write in a polished, human-like narrative style, as if prepared by a senior analyst.
- Keep the report concise yet detailed, with clear explanations.
- Reference plots by filename including the directory (e.g., '{output_dir}/plot_query_1.png') without describing their content.
- If results are empty, note potential causes (e.g., no data, query error) and provide insights accordingly.
- Return the report as plain text, without markdown, code fences, or additional comments.

\end{lstlisting}

\subsubsection{CHART\_PROMPT}
\begin{lstlisting}
Generate Python code that:
Starts with necessary imports: import matplotlib.pyplot as plt and from decimal import Decimal. Include import matplotlib.dates as mdates only if plotting time-series data with dates.
Uses the provided data variable directly (e.g., data = {data}), without hardcoding data or including placeholder comments like # Add all data here....
Converts Decimal values to floats for plotting (e.g., float(x) for values like 'job_count' or 'completion_rate').
Handles null or missing values by replacing them with 0 or skipping them using try/except.
Creates a comparative plot using plt.figure(), plt.bar() or plt.plot(), plt.title(), plt.xlabel(), plt.ylabel(), plt.grid(True), plt.tight_layout(), plt.savefig(), and plt.close().

Adapts to the query intent:
For categorical data (e.g., job states, lab_id), use plt.bar() with the first column as x-axis labels and the second as y-axis values (e.g., counts).
For time-series data (e.g., daily counts), use plt.plot() or plt.plot_date() with the first column (dates) as x-axis and subsequent columns as y-values. Use mdates for date formatting if needed.
For dual metrics (e.g., counts and averages), use a dual-axis plot with plt.twinx().
For sparse or mixed data (e.g., hours and states), create a bar or scatter plot, grouping by the first column.
For large categorical datasets (e.g., many unique IDs), limit to the top 10 categories by value, sorted descending.
For non-plottable data, generate a placeholder plot with a descriptive message.
Saves the plot as '{plot_filename}' in the specified directory.
Ensures all strings are properly terminated with matching quotes and code is syntactically correct.
Returns only the Python code as plain text, without comments, explanations, markdown, code fences, or any text like 'Here's the code:'.

Example for time-series data:

import matplotlib.pyplot as plt
import matplotlib.dates as mdates
from decimal import Decimal
data = {data}
dates = [x['date_day'] for x in data]
counts = [float(x['job_count']) if x['job_count'] is not None else 0 for x in data]
plt.figure(figsize=(10, 6))
plt.plot(dates, counts, 'b-', label='Job Count')
plt.xlabel('Date')
plt.ylabel('Job Count')
plt.title('Daily Job Counts')
plt.grid(True)
plt.xticks(rotation=45)
plt.tight_layout()
plt.savefig('{plot_filename}')
plt.close()

Example for categorical data:

import matplotlib.pyplot as plt
from decimal import Decimal
data = {data}
labels = [x['lab_id'] for x in data]
values = [float(x['job_count']) if x['job_count'] is not None else 0 for x in data]
sorted_data = sorted(zip(labels, values), key=lambda x: x[1], reverse=True)[:10]
labels, values = zip(*sorted_data)
plt.figure(figsize=(10, 6))
plt.bar(labels, values)
plt.title('Top 10 Lab IDs by Job Count')
plt.xlabel('Lab ID')
plt.ylabel('Job Count')
plt.xticks(rotation=45)
plt.grid(True, axis='y')
plt.tight_layout()
plt.savefig('{plot_filename}')
plt.close()

Example for dual-axis time-series:

import matplotlib.pyplot as plt
import matplotlib.dates as mdates
from decimal import Decimal
data = {data}
dates = [x['date_day'] for x in data]
counts = [float(x['job_count']) if x['job_count'] is not None else 0 for x in data]
avgs = [float(x['avg_time']) if x['avg_time'] is not None else 0 for x in data]
fig, ax1 = plt.subplots(figsize=(10, 6))
ax1.plot(dates, counts, 'b-', label='Job Count')
ax1.set_xlabel('Date')
ax1.set_ylabel('Job Count', color='b')
ax1.tick_params(axis='y', labelcolor='b')
ax2 = ax1.twinx()
ax2.plot(dates, avgs, 'r-', label='Avg Time (s)')
ax2.set_ylabel('Avg Time (s)', color='r')
ax2.tick_params(axis='y', labelcolor='r')
plt.title('Daily Job Counts and Avg Execution Time')
ax1.grid(True)
plt.xticks(rotation=45)
plt.tight_layout()
plt.savefig('{plot_filename}')
plt.close()

Example for empty data:

import matplotlib.pyplot as plt
plt.figure(figsize=(8, 6))
plt.text(0.5, 0.5, 'Data Unavailable', ha='center', va='center', fontsize=12)
plt.title('Query Results')
plt.axis('off')
plt.tight_layout()
plt.savefig('{plot_filename}')
plt.close()
\end{lstlisting}

\subsection*{A2. Additional Output}
This section presents a detailed example from the Cycle Time Reduction Agents (CTRA) workflow, illustrating a successful query execution to demonstrate the collaborative functionality of the LangGraph-based system. The example analyzes the average time from job creation to start, grouped by job state, using a dataset of 5,031 records from a PostgreSQL 'jobs' table, representing one month of laboratory operations in a pharmaceutical and biotechnology context. The roles of the \textbf{Question Creation Agent}, \textbf{Query Builder}, \textbf{Query Validator}, \textbf{Summarization Agent}, and \textbf{Charting Agent} are delineated, with human-actionable suggestions emphasized to facilitate laboratory optimization.

\paragraph{Example: Average Creation-to-Start Time by Job State}
This example examines delays between job creation and initiation, a critical metric for identifying inefficiencies in scheduling or resource allocation within laboratory operations. By grouping results by job state, the analysis elucidates how different job statuses (e.g., completed, running, or paused) influence cycle times, offering insights into potential bottlenecks.

\begin{itemize}
  \item \textbf{Question Creation Agent}: Generated the question: ``What is the average creation-to-start time for jobs, grouped by state? (Suitable for bar chart).'' This question addresses delays in job initiation, a pivotal factor in cycle time reduction, as such delays may reflect issues in scheduling, equipment availability, or personnel preparedness. The agent selected a bar chart as the optimal visualization to compare average times across job states, enabling laboratory managers to identify states with excessive delays efficiently.

  \item \textbf{Query Builder}: Constructed the following SQL query:
\begin{lstlisting}
SELECT state, AVG(EXTRACT(EPOCH FROM (started_timestamp - created_timestamp))) AS avg_creation_to_start_time
FROM jobs
WHERE started_timestamp IS NOT NULL
GROUP BY state
\end{lstlisting}

  \item \textbf{Query Validator}: Verified the query's correctness and suggested indexing on 'started\_timestamp' and 'state' for performance.

  \item \textbf{Query Results}:
  \begin{itemize}
    \item COMPLETED: 8,693.34 seconds (\~145 minutes)
    \item UNSCHEDULED: 5,991.09 seconds (\~100 minutes)
    \item CANCELLED: 3,398.24 seconds (\~57 minutes)
    \item RUNNING: 3,486.02 seconds (\~58 minutes)
    \item IN\_ERROR: 41.50 seconds (\~0.7 minutes)
    \item PAUSED: 33.13 seconds (\~0.6 minutes)
  \end{itemize}

  \item \textbf{Summarization Agent}:
  \begin{itemize}
    \item Identified high delays for COMPLETED and UNSCHEDULED.
    \item Suggested causes: scheduling inefficiencies, equipment unavailability, or personnel delays.
    \item Noted rapid but possibly error-prone initiation for IN\_ERROR and PAUSED jobs.
    \item Highlighted balanced but notable delays in CANCELLED and RUNNING states.
  \end{itemize}

  \item \textbf{Human-Actionable Suggestions}:
  \begin{itemize}
    \item Review and optimize scheduling protocols.
    \item Investigate equipment availability.
    \item Standardize setup protocols across labs.
    \item Analyze reasons for job cancellations.
    \item Conduct root cause analysis for IN\_ERROR jobs and enhance personnel training.
  \end{itemize}

  \item \textbf{Charting Agent}: Produced a bar chart saved as Figure \ref{fig:fig3}.
  \begin{itemize}
    \item X-axis: Job states
    \item Y-axis: Avg creation-to-start time (seconds, log scale)
    \item Distinct colors and data labels for clarity
    \item Title: \textit{Average Creation-to-Start Time by Job State}
  \end{itemize}
\end{itemize}

This example underscores the CTRA workflow's capability to systematically analyze laboratory data, identify inefficiencies, and provide actionable insights for improving pharmaceutical and biotechnology operations.

\end{document}